\documentstyle[11pt,paspconf,psfig]{article}

\input epsf.sty 
\begin{document}

\title{On the Period-Luminosity-Color Relation of Classical Cepheids}

\author{G. Bono\altaffilmark{1}}
\affil{Osservatorio Astronomico di Roma, Via Frascati 33, 00040 Monteporzio, 
Italy, {\tt bono@coma.mporzio.astro.it}}

\author{M. Marconi\altaffilmark{2}}
\affil{Osservatorio Astronomico di Napoli, Via Moiariello 16, 80100 Napoli, 
Italy, {\tt marcella@na.astro.it}}

\altaffiltext{1}{On leave from Osservatorio Astronomico di Trieste, 
Via G.B. Tiepolo 11, 34131 Trieste, Italy} 
\altaffiltext{2}{Dipartimento di Fisica Universit\`a di Pisa, Piazza 
Torricelli 2, 56126 Pisa, Italy} 

\begin{abstract}
We present the comparison between theoretical and empirical PL relations 
in V and I bands for the Large Magellanic Cloud. We found that, within the 
current intrinsic dispersions, theoretical predictions are in remarkable 
agreement with observational data. We also discuss the PLC relations in 
(V,B-V) and (V,V-I) as well as the dependence of distance determinations 
on uncertainties in colors, in reddening corrections and in metal content.  
\end{abstract}



\section{Introduction}

Classical Cepheids are the most popular primary distance indicators. The 
key role played by this group of radial variables in estimating the 
cosmic distance scale is soundly confirmed by the large number of both 
observational and theoretical investigations aimed at improving both  
accuracy and reliability of distance measurements. 
At least three are the main reasons for the widespread use of the Cepheid 
distance scale:  

$\bullet$ Cepheids are bright intermediate-mass objects whith visual 
magnitude range from $M_V$=-3 at short-periods ($logP\approx0.6$) to $M_V$=-6 
at long-periods ($logP\approx2.0$). This feature and the luminosity 
variations over the pulsation cycle make these variables an appealing 
observational target, since they can be detected and measured in a large 
number of Local Group (LG) galaxies. This observational effort has reached 
the top thanks to the photometric data recently collected by the HST.  
In fact, the HST key-project succeeded in the identification of a 
reasonable number of Cepheids in two dozens galaxies belonging to the 
LG and to the Virgo cluster. 

$\bullet$ The physical mechanisms which govern the pulsation instability 
of these objects have been firmly established. This notwithstanding, 
the pulsation 
behavior and the modal stability of Cepheids are often outlined by 
adopting n\"aive arguments. Moreover, up to now no general theoretical 
consensus on the dependence of the Cepheid luminosity on the metal abundance 
has been reached. In fact, theoretical predictions based on linear, 
nonadiabatic, radiative models suggest that the blue edge of the 
instability strip presents a negligible dependence on metal content 
(Saio \& Gautschy 1998). On the other hand, recent theoretical predictions 
which account for both blue and red edges of the instability strip support 
the evidence that Cepheid luminosity depends on  chemical composition. 
In particular, it turns out that metal-poor Cepheids are, at fixed 
period, brighter than metal-rich ones. This prediction is at odds with 
their empirical behavior, and indeed a trend opposite to this has been recently 
brought out in the literature (Sasselov et al. 1997; Kennicutt et al. 1998 
and references therein).  

$\bullet$ Some of the long-standing questions on distance determinations 
can be properly addressed within the Cepheid scenario. In fact, the 
intrinsic width of the instability strip substantially narrows when moving 
from optical to near infrared (NIR) bands. Metallicity dependence and 
reddening corrections give the same outcome, and indeed NIR 
PL relations are less affected by these uncertainties than optical PL 
relations. The calibration of the PL zero-points can be accomplished by 
adopting several independent methods such as the 
Baade-Wesselink method and its progeny (Krockenberger et al. 1997; 
Di Benedetto 1997), the trigonometric parallaxes (Feast \& Catchpole 1997),  
and the main sequence fitting (Gieren \& Fouqu\`e 1993 and references 
therein). Therefore a straightforward analysis of both systematic and 
intrinsic errors affecting distance determinations may be undertaken. 

Even though Cepheids are the most promising primary distance indicators
for estimating extragalactic distances, the present scenario of
the cosmic distance scale, and in turn of the evaluation of the Hubble 
constant -$H_0$- has been perfectly accounted for by Trimble (1997) 
who noted that although over the years $H_0$ estimates decreased by one order 
of magnitude, the error bars have remained almost constant!   
This evidence strongly supports a comprehensive theoretical 
investigation of the deceptive errors which could affect distance 
measurements.

In section 2 we discuss the comparison between theoretical and empirical
PL relations in V and I bands for Cepheids in the Large Magellanic Cloud 
(LMC), as well as the so 
called Wesenheit function. Theoretical prediction concerning the 
Period-Luminosity-Color (PLC) relations are presented in \S 3 together 
with a brief analysis of the uncertainties due to both metallicity and 
reddening.

\section{PL relations for LMC Cepheids}

LMC Cepheids are the backbone of cosmic distance derivation since 
the PL relations of the target galaxy are compared with the LMC PL 
relations when placing extragalactic distances on an absolute scale.
Even though the Cepheid metallicity is not firmly 
constrained (see Luck et al. 1998) and the reddening estimates 
still present some uncertainties, LMC Cepheids cover a wide period 
range and are relatively close objects, thus enabling a proper 
sampling of the instability strip. 

In a recent paper Tanvir (1997) performed a thorough analysis of 
the intrinsic and systematic uncertainties which affect Cepheid 
distance scale. In particular, the author derived new LMC PL 
relations based on current available data in V and I bands, as well 
as a PL relation for the reddening free Wesenheit function, i.e. 
$W_{VI}=V -R[V-I]$ where $R\approx2.45$ is the adopted extinction 
parameter (Cardelli et al. 1989). 
These empirical relations were calibrated by adopting an LMC distance 
modulus of 18.5 and a reddening of $E_{B-V}=0.1$ mag, respectively. 
In order to supply a theoretical framework for a proper
comparison with observational data we derived analytical PL relations 
by adopting several sequences of Cepheid models constructed with  
fixed chemical composition (Y=0.25 Z=0.008) and a wide range 
of stellar masses and effective temperatures (see Bono et al. 1999a,b 
for further details). The mass-luminosity relation adopted for fixing 
the luminosity of these models is based on evolutionary tracks which 
neglect the convective core overshooting during the hydrogen burning 
phase. 

In deriving these relations we also nailed on the period at $log P=1.4$,
but we did not restrict the period range to $log P<1.8$ (Tanvir 1997), 
since we are interested in testing theoretical predictions for 
long-period Cepheids. Due to the well known bending of the PL relation 
in the long-period range we performed a quadratic fit and the results 
we obtained are the following:

\begin{displaymath}
\small 
\begin{array}{cllll}
<M_V> =&-5.18&-2.01\; [log P-1.4]&+ 0.88\; [log P-1.4]^2&\;(\sigma_{rms}=0.25)\\
<M_I> =&-6.09&-2.43 [log P-1.4]&+ 0.67 [log P-1.4]^2&\;(\sigma_{rms}=0.18)\\
<W_{VI}>=&-7.41&-3.02\;[log P-1.4]&+ 0.38\;[log P-1.4]^2&\;(\sigma_{rms}=0.08)\\
\end{array}
\end{displaymath}

Figure 1 shows the comparison between theoretical and empirical PL 
relations. From this comparison  three interesting results emerge:
1)  In the period range covered 
by empirical relations ($0.4<log P<1.7$) the mean $PL_V$ and $PL_I$ relations
are, within the intrinsic dispersions, in remarkable agreement 
with theoretical predictions. 
2) The theoretical PL relations show a linear behavior up to 
$logP\approx1.6$ but toward longer periods they start to bend due to 
the shift of the instability strip toward redder colors. 
3) Theoretical and empirical dispersions in the I band are, as expected, 
smaller than those of the V band,  but they are still too large for 
constraining the 
distance modulus and the reddening values adopted for calibrating 
this sample. This notwithstanding, there is no plausible reason for 
assuming that the pulsation characteristics of LMC Cepheids are peculiar
(Simon \& Young 1997), thus supporting the use of these templates 
for estimating the absolute distance of target galaxies with similar 
metal contents. 

\begin{figure}
\vbox{
\hbox{
\centerline{\psfig{file=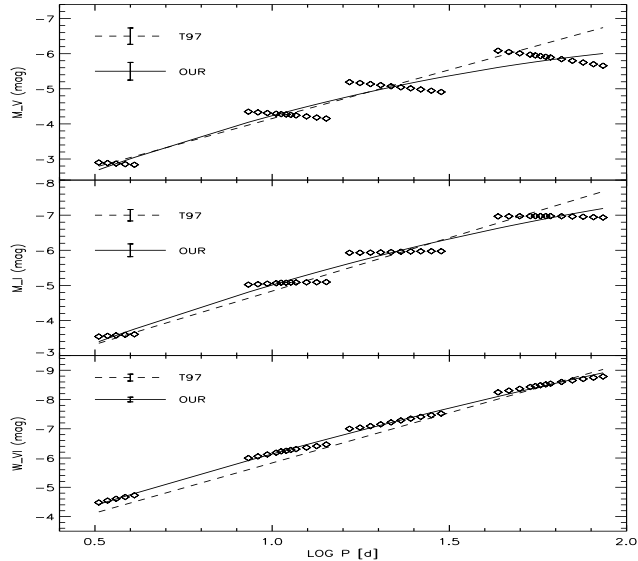,height=7cm,width=9.0cm}}
}}

\caption{Comparison between our theoretical PL relations (solid lines)
and the empirical relations (dashed lines) provided by Tanvir (1997). 
The error bars are referred to the intrinsic dispersions of analytical 
relations. From top to bottom the relation plotted in this figure are 
the $PL_V$, the $PL_I$ and the Wesenheit function.} 
\end{figure}

The bottom panel of Figure 1 shows the comparison between theoretical 
and empirical $PL_W$ relations based on Wesenheit magnitudes. 
The two relations are in agreement only marginally, and when moving 
from long to short-periods the discrepancy increases. Due to the 
agreement between theory and observations in the $logP-M_V$ and in the 
$logP-M_I$ plane, it is not clear whether this drift is caused by a 
systematic shift in the color-temperature relations we adopted or 
more likely is caused by a poor accuracy of (V-I) mean colors. 
In fact, in order to reduce the sampling of I light curves the V band 
light curves are transformed into I band light curves by adopting an 
empirical method (see Appendix A in Tanvir 1997). We note
that the light curves are not characterized by a constant shape when 
moving from short to long-period Cepheids. In the short-period range 
they present a sawtoothed shape, then around $logP\approx1$ they show 
a bump along either the rising or the decreasing branch while in the 
long-period range the shape becomes more sinusoidal. 

It has been often pointed out that the use of these magnitudes causes a 
substantial decrease in the dispersion of the PL relation. This is mainly 
due to the fact that the Wesenheit function in two arbitrary bands 
($\beta$, $\xi$) is the projection of the PLC ($\beta$, $\beta-\xi$) 
relation onto the $logP-M_\beta$ plane. As a consequence, in this plane
the dispersion  is much smaller since the location of each 
individual Cepheid inside the instability strip is being fixed according to 
both period and color. 

{\em Is therefore the Wesenheit function equivalent to using a PLC 
relation for estimating distances?}   

On general grounds the answer is no, because the Wesenheit function is 
only mimicking the behavior of the equivalent PLC relation. Due to an  
unaccountable reason the value of the color coefficient in the 
optical PLC (V, B-V) relation is quite similar (see Bono et al. 1999b) 
to the extinction parameter,
and thus in these bands the Wesenheit function and the PLC relation 
present the same behavior. However, the Wesenheit function (V,I) is 
derived by adopting $R\approx2.45$ (Cardelli et al. 1989) and therefore
it is quite different from the color coefficient of the PLC (V, V-I) 
relation (see {\em infra}). 

On the basis of the above results three main conclusions  
concerning the use of the $logP-W_{VI}$ relation can be drawn:
1) in these bands the Wesenheit function does not mimic the behavior 
of the PLC (V, V-I) relation and therefore it does not properly account 
for the intrinsic width of the instability strip.  
2) This relation is sensitive to the photometric accuracy of mean colors 
and therefore for a sound empirical evaluation a good 
sampling of both V and I band light curves is needed. 
This means that if accurate mean colors are available the use of the 
PLC (V, V-I) relation for estimating distances is more physically 
plausible than the Wesenheit function. The interesting 
feature that the Wesenheit function supplies reddening free magnitudes
will be discussed in a forthcoming paper (Caputo et al. 1999).

\section{Discussion and Conclusions} 

The above discussion has been focused on the PL relations at fixed 
chemical composition. In order to account for the dependence of 
distance determinations on metallicity we adopt the PLC relations, 
since in this parameter space the Cepheid location is not ambiguous.
We derived two analytical relations by taking into account sequences 
of models constructed by adopting three 
different chemical compositions, namely Y=0.25, Z=0.004/[Fe/H]=-0.7; 
Y=0.25, Z=0.008/[Fe/H]=-0.4 Y=0.25, Z=0.02/[Fe/H]=0.0. 
The results of this fit for (V, B-V) and (V, V-I) PLC relations are 
the following: 

\begin{displaymath}
\begin{array}{lllll}
<M_V> =&-2.83   &- 3.57 log P &+ 2.92 [<B>-<V>] &- 0.35 [Fe/H] \\
       &\pm 0.05&\pm 0.03     &\pm 0.05         &\pm 0.03      \\ 
       &        &             &                 &              \\ 
<M_V> =&-3.57   &- 3.59 log P &+ 3.80 [<V>-<I>] &+ 0.03 [Fe/H] \\ 
       &\pm 0.03&\pm 0.02     &\pm 0.04         &\pm 0.02                 
\end{array}
\end{displaymath}

where the symbols have their usual meaning. The standard deviations of these 
two relations are $\sigma_{rms}=0.05$ and 0.03 respectively. 
As already pointed out by Bono et al. (1999b) the PLC (V, B-V) relation 
suggests that metal-rich Cepheids are, at fixed period and color, brighter 
than metal-poor ones. On the other hand, the PLC (V, V-I) relation  
discloses a marginal dependence on metallicity, and thus leads strong 
support to the use of this relation for estimating distances of 
target galaxies for which the metal content is poorly known. 

Since these relations are not affected by systematic errors such as 
reddening corrections, photometric accuracy of both magnitude and 
colors, and the metal content, we can estimate how plausible errors 
within these parameters may affect distance determinations. 
At first, to account for photometric and zero-point calibration 
uncertainties we assumed, by following Tanvir (1997), that 
$\sigma_B\approx\sigma_V\approx\sigma_I\approx0.04$ mag. These 
uncertainties imply errors on (B-V) and (V-I) colors of the order 
of 0.06 mag, and in turn an uncertainty in distance of 8\% and 10\%. 
At the same time, a systematic error of the order of 0.03 mag in 
the reddening correction -$E_{B-V}$- implies errors on (B-V) and (V-I) 
colors of 0.03 and 0.04 mag respectively. These uncertainties imply errors 
in distance equal to 4\% and 7\%. However, if we simultaneously account 
for the two quoted uncertainties, the errors on the colors are equal 
to 0.19 and 0.26 mag, while in distances the errors are 9\% and 12\% 
respectively. 

In order to account for the dependence of distance determinations
on chemical composition we assumed an uncertainty on metal abundance 
of the order of 0.4 dex. This error takes into account not 
only the intrinsic errors in metallicity measurements but also the 
spread in metallicity of Cepheids in external galaxies. In fact, 
Monteverde et al. (1997) by adopting spectroscopic data of B-type 
supergiants in M33 estimated that the iron abundance gradient in 
this spiral galaxy is of the order $-0.20(\pm 0.05)$ dex kpc$^{-1}$  
The quoted uncertainty implies errors in distance modulus of 
0.14 (V,B-V) and of 0.012 (V,V-I) mag and thus errors of 7\% and 
1\% respectively.
 
The main outcomes of this leading term error analysis are the 
following: 
1) PLC relations based on different photometric bands present pros and 
cons, and indeed the PLC (V,B-V) relation, in comparison with the 
PLC (V,V-I) relation, is less affected by errors on both reddening 
corrections and colors but is more affected by a spread in metallicity.  
2) The PLC (V,V-I) relation seems a promising distance indicator for 
target galaxies with accurate reddening and color measurements since 
it is marginally affected by metallicity.  

Obviously a comprehensive analysis of the error budget of Cepheid 
distance scale based on PLC relations is not a trivial effort, since 
in reality it is a mixture of all previous uncertainties. A thorough analysis
can be undertaken only by means of Montecarlo simulations of Cepheids 
inside the instability strip, which can simultaneously account for all 
plausible errors on observables.  

%
%
\acknowledgments
This work was partially supported by CNA through a postdoc research 
grant to M. Marconi, by ASI and CRA.


\end{document}